\documentclass[aps,prd,10pt,amsmath,amsfonts,twocolumn,nofootinbib]{revtex4-1}

\usepackage[utf8]{inputenc}

\usepackage{bm}
\usepackage{graphicx}
\usepackage{physics}

\usepackage{color}

\usepackage{physics}
\usepackage[colorlinks]{hyperref}

\hypersetup{
	bookmarksnumbered,
	pdfstartview={FitH},
	citecolor={green},
	linkcolor={darkblue},
	urlcolor={darkblue},
	pdfpagemode={UseOutlines}}
\definecolor{darkgreen}{RGB}{20,100,20}
\definecolor{darkblue}{RGB}{0,0,130}
\definecolor{darkred}{rgb}{.8,0,0}

\newcommand{\psii}{\psi_\text{I}}
\newcommand{\psiii}{\psi_\text{II}}
\newcommand{\phii}{\phi_\text{I}}
\newcommand{\phiii}{\phi_\text{II}}
\newcommand{\ali}{\alpha_\text{I}}
\newcommand{\alii}{\alpha_\text{II}}
\newcommand{\bei}{\beta_\text{I}}
\newcommand{\beii}{\beta_\text{II}}
\newcommand{\ii}{\text{I}}
\newcommand{\iii}{\text{II}}

\newcommand{\A}{\mathcal{A}}
\newcommand{\Ai}{\mathcal{A}_\text{I}}
\newcommand{\Aii}{\mathcal{A}_\text{II}}
\newcommand{\La}{\Lambda}
\newcommand{\aL}{\alpha_{\Lambda}}
\newcommand{\bL}{\beta_{\Lambda}}
\newcommand{\psiL}{\psi_{\Lambda}}
\newcommand{\phiL}{\phi_{\Lambda}}
\newcommand{\AL}{\mathcal{A}_{\Lambda}}

\begin{document}

\title{Effect of relativistic acceleration on continuous variable quantum teleportation and dense coding}

\author{Piotr T. Grochowski}
\email{pg347902@okwf.fuw.edu.pl}
\author{Grzegorz Rajchel}
\email{grzegorz.rajchel@student.uw.edu.pl}
\author{Filip Kiałka}
\email{fk322204@okwf.fuw.edu.pl}
\altaffiliation{Present address: University of Duisburg-Essen, Lotharstraße 1-21, 47057 Duisburg, Germany.}
\author{Andrzej Dragan}
\email{dragan@fuw.edu.pl}

\affiliation{Institute of Theoretical Physics, Faculty of Physics, University of Warsaw, Pasteura 5, 02-093 Warsaw, Poland}

\date{\today}

\begin{abstract}
	We investigate how relativistic acceleration of the observers can affect the performance of the quantum teleportation and dense coding for continuous variable states of localized wavepackets.
	Such protocols are typically optimized for symmetric resources prepared in an inertial frame of reference.
	A mismatch of the sender's and the receiver's accelerations can introduce asymmetry to the shared entanglement, which has an effect on the efficiency of the protocol that goes beyond entanglement degradation due to acceleration.
	We show how these asymmetry losses can be reduced by an extra LOCC step in the protocols.
\end{abstract}

\maketitle

\section{Introduction}
	The discovery of the Unruh effect~\cite{Fulling1973,Davies1975,Unruh1976} revealed that the vacuum state of any quantum field defined by an inertial observer appears to be thermal according to a uniformly accelerated observer.
	This has led to a surprising conclusion that all quantum states are in general observer-dependent.
	This intriguing property of relativistic quantum fields has a striking impact on the theory of quantum information in which the notion of quantum states plays a crucial role.
	The conclusion that the effect of motion on quantum states can in principle affect all types of quantum information protocols between moving parties has led to a growing interest in a new field of research -- relativistic quantum information.
	Since at the heart of many of these protocols lies a crucial ingredient -- quantum entanglement, many efforts have been undertaken to study how it is affected by relativistic acceleration or gravity treated as a classical background for the quantum fields.
	The first works on the topic~\cite{Alsing2003,Fuentes2005} have studied how uniformly accelerated motion can lead to the reduction of entanglement between two field modes shared by a pair of observers in relative motion.
	Oversimplifications of the approach used by these authors have soon been pointed out and resulted in a more refined approach going beyond the so-called single-mode approximation~\cite{Bruschi2010}.
	Unfortunately, also this approach followed by many authors~\cite{papers} failed to provide a physically satisfactory interpretation of the results of the calculations due to unclear character of the global modes used in the setup~\cite{Dragan2013a}.
	Two possible routes overcoming these difficulties have been proposed as a way out.
	Both of them rely on a replacement of global Unruh modes used in the description of quantum states by localized quantum states.
	In the first approach one introduces an ideal cavity that can store and transport quantum states along an arbitrary path~\cite{Bruschi2012,Bruschi2013}, the other approach involves using approximately localized wave-packets that are stationary either in an inertial, or in a uniformly accelerated frame of reference~\cite{Dragan2013a,Dragan2013,Doukas2013,Ahmadi2016}.
	
	One has to keep in mind that entanglement is eventually only a resource for communication protocols such as quantum teleportation~\cite{Bennett1993} or dense coding~\cite{Bennett1992} and therefore it is important to take into account the effect of acceleration on the whole protocol and not only on one of its ingredients.
	In particular, a teleportation protocol is typically optimized for symmetric settings, but a non-inertial motion of the observers can introduce asymmetry into the shared entanglement that can also have impact going beyond entanglement degradation of the resource alone.
	In this work we fill in the gap in the existing literature by considering the two aforementioned protocols: quantum teleportation and dense coding of continuous variables taking into account the effect of an arbitrary accelerated motion of two independent parties: the sender and the receiver.
	We will consider the most general scenario using the localized framework for Gaussian states introduced in Ref.~\cite{Ahmadi2016}, in which both parties can move with independent arbitrary relativistic accelerations and be separated by an arbitrary distance.
	Finally, we will show how the efficiency losses due to observers' unequal accelerations can be reduced by performing a motion-dependent, noisy LOCC operation before the measurements.

	The work is organized in the following way.
	In Sec.~\ref{sec:effect_of_acceleration_on_states} we analyze the effect of the uniform acceleration on a two-mode Gaussian state which is utilized later as the resource for the quantum-information protocols.
	We introduce the measures of efficiency for given protocols and calculate them for the accelerated parties in the following sections.
	Sec.~\ref{sec:continuous_variable_quantum_teleportation} and Sec.~\ref{sec:continuous_variable_dense_coding} are devoted to quantum teleportation and dense coding, consecutively.
	In Sec.~\ref{sec:locc_optimization} we present the LOCC optimization strategies for both protocols.
	Finally, Sec.~\ref{sec:conclusions} provides conclusions and outlook.

\section{Effect of relativistic acceleration on Gaussian states\label{sec:effect_of_acceleration_on_states}} 

	\begin{figure}[t]
		\centering
		\includegraphics[width=0.48\textwidth]{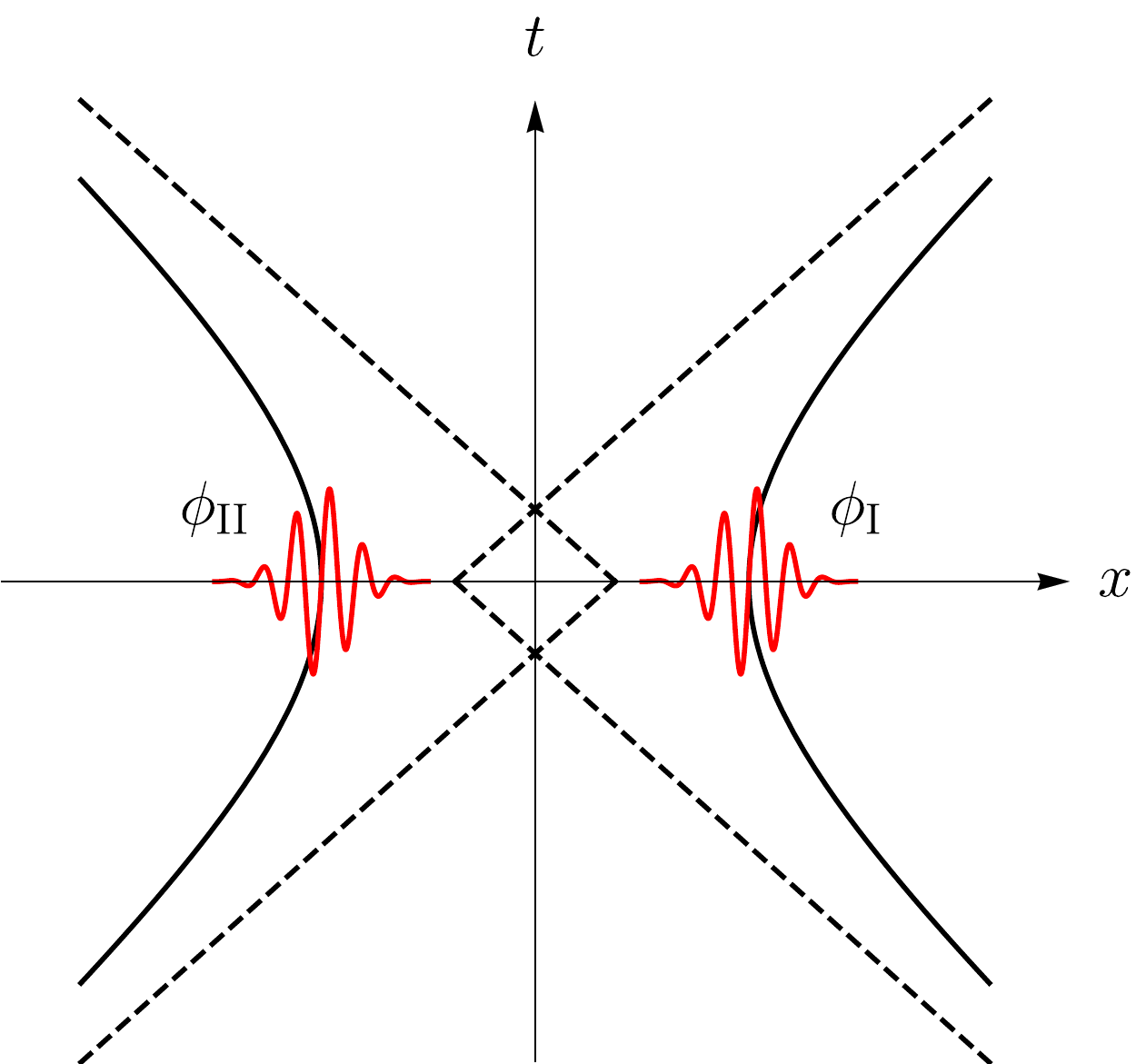}
		\caption{The trajectories of accelerating parties Alice (right) and Bob (left) who perform a quantum-information protocol using the state of inertial wavepackets $\phii$ and $\phiii$ as a resource.
		The two Rindler wedges intersect to allow for classical communication.
		The wavepackets are drawn outside the intersection region for clarity.
		\label{rind}}
	\end{figure}

	We consider two uniformly accelerating observers which perform certain quantum-information protocol using a two-mode state prepared in an inertial frame (see Fig.~\ref{rind}).
	To transform this state from the inertial frame to the rest frame of the observers, we employ a Bogolyubov transformation between suitably chosen bases of mode functions~\cite{Fuentes2005}.
	The initial basis includes, in particular, the spatially localized mode functions in which the resource state is prepared.
	Ref.~\cite{Ahmadi2016} presents a method that allows to perform such Bogolyubov transformation for an arbitrary two-mode Gaussian state.
	Moreover, the approach of Ref.~\cite{Ahmadi2016} is not constrained by the standard geometry of the Rindler chart.
	That is, it allows to independently tune the proper accelerations of the observers and the minimal distance between them.
	In this section we review the basic elements of this approach, apply it to a two-mode squeezed vacuum state, and simplify by using certain approximations.

	\subsection{Effect of acceleration as the action of a quantum channel\label{sub:effect_of_acceleration_as_the_action_of_a_quantum_channel}} 

	From now on we specialize to 1+1-dimensional flat spacetime and adopt units in which $c=\hbar=1$.
	Let us consider a real, scalar, massive quantum field, which satisfies the Klein-Gordon equation
	\begin{equation} \label{eq:KG}
	 	\left(\Box + m^2\right)\hat{\Phi}=0.
	 \end{equation}
	The field operator $\hat{\Phi}$ can be decomposed using an orthonormal basis of solutions of Eq.~\eqref{eq:KG}.
	We will consider two such decompositions: one corresponding to an inertial Minkowski observer and one corresponding to the uniformly accelerating Rindler observers.
	The first decomposition consists of mode functions $\phi_k$ that contain only positive frequencies with respect to the Minkowski timelike Killing vector.
	Analogously, the second decomposition consists of mode functions $\psi_k$ that contain only positive frequencies with respect to the Rindler timelike Killing vector.
	We will denote the annihilation operators associated with the two decompositions by $\hat{f}_k$ and $\hat{d}_k$, respectively.
	The field operator can therefore be written as
	\begin{equation}
		\hat{\Phi}=\sum_k \phi_k \hat{f}_k + H.c. = \sum_k \psi_k \hat{d}_k + H.c.
	\end{equation}

	We now assume that only two of the $\phi_k$ modes are occupied in the initial state.
	We will denote them by $\phii$, $\phiii$ and the corresponding annihilation operators by $\hat{f}_{\text{I}}$, $\hat{f}_{\text{II}}$.
	Moreover, we will restrict ourselves to only two of the $\psi_k$: $\psii$ and $\psiii$, with the annihilation operators $\hat{d}_{\text{I}}$ and $\hat{d}_{\text{II}}$.
	The remaining $\psi_k$ are not empty, but we assume that the accelerating observers only have access to one mode each.

	It was shown in Ref.~\cite{Holevo2001} that the transformation of the state from one frame of reference to another is a noisy Gaussian channel.
	This means that for a Gaussian input state the output state is also Gaussian.
	To write down the action of this channel, we first introduce the quadrature operators associated with $\phiL$, $\La \in \lbrace \text{I},\text{II} \rbrace$:
	\begin{align}
		\hat{q}_{\La}^{(f)} = \frac{\hat{f}_{\La} + \hat{f}_{\La}^{\dagger}}{\sqrt{2}},\qquad
		\hat{p}_{\La}^{(f)} = \mathrm{i}\frac{\hat{f}_{\La}^{\dagger} - \hat{f}_{\La}}{\sqrt{2}}.
	\end{align}
	The quadratures associated with $\psiL$, which we denote by $\hat{q}_{\La}^{(d)}$ and $\hat{p}_{\La}^{(d)}$, are defined analogously.
	We then gather the relevant quadratures into a vector
	\begin{equation}
		\hat{\bm{X}}^{(i)}=\left(\hat{q}_{\text{I}}^{(i)},\hat{p}_{\text{I}}^{(i)},\hat{q}_{\text{II}}^{(i)},\hat{p}_{\text{II}}^{(i)}\right)^T,
	\end{equation}
	where $i\in\{f,d\}$.
	With these, the first statistical moments of the state are written as the expectation values
	\begin{equation}
		\bm{X}^{(i)}=\expval{\hat{\bm{X}}^{(i)}},
	\end{equation}
	and the second moments are given by a covariance matrix
	\begin{equation}
		\sigma_{kl}^{(i)}=\frac{1}{2}\left\langle \left\lbrace \hat{X}_k^{(i)},\hat{X}_l^{(i)}   \right\rbrace\right\rangle-\left\langle \hat{X}_l^{(i)}\right\rangle\left\langle \hat{X}_k^{(i)}\right\rangle.
	\end{equation}
	Finally, the Bogolyubov transformation we are interested in can then be written as~\cite{Holevo2001}:
	\begin{subequations} \label{eq:channel}
	\begin{align}
			\bm{X}^{(d)} &= M\bm{X}^{(f)},\\
			\sigma^{(d)} &= M\sigma^{(f)}M^T+N,
	\end{align}
	\end{subequations}
	where $N$ is the noise matrix~\cite{Ahmadi2016} and $M$ is given in terms of overlaps of $\phiL$ with $\psiL$.

	If we define
	\begin{subequations} \label{eq:overlaps}
	\begin{align}
			\aL &= (\psiL,\phiL),\\
			\bL &= -(\psiL,\phiL^\star),
	\end{align}
	\end{subequations}
	where $(\cdot,\cdot)$ denotes the Klein-Gordon scalar product, then the $M$ matrix is~\cite{Ahmadi2016}:
	\begin{widetext}
	\begin{equation} \label{eq:M}
		M =
		\begin{pmatrix}
		\Re(\ali-\bei) & -\Im(\ali+\bei) & 0 & 0 \\
		\Im(\ali-\bei) & \Re(\ali+\bei) & 0 & 0 \\
		0 & 0 & \Re(\alii-\beii) & -\Im(\alii+\beii) \\
		0 & 0 & \Im(\alii-\beii) & \Re(\alii+\beii)
		\end{pmatrix}.
	\end{equation}
	\end{widetext}
	We have found, however, that in the cases we study in this paper, the $\bL$ coefficients are negligibly small compared to $\aL$\footnote{
		\label{foot:neglecting_betas}
		The calculated values of $\bL$ were at least 8 orders of magnitude smaller than the values of $\aL$.
		We note that neglecting $\bL$ implies that the channel does not depend on the minimal distance between the observers.
		Provided the choice of mode functions is fixed, the only free parameters are the observers' proper accelerations.
	}. If we omit them, the $M$ matrix simplifies to
	\begin{subequations}  \label{eq:MN_matrices}
	\begin{equation}
		M = \ali\openone \oplus \alii\openone,
	\end{equation}
	where $\openone$ is a $2\times2$ identity matrix.
	The $N$ matrix, on the other hand, is then given by
	\begin{equation}
		N = \left( 1-\ali^2 \right) \openone \oplus \left( 1-\alii^2 \right) \openone.
	\end{equation}
	\end{subequations}

	\subsection{Choice of the mode functions\label{sub:choice_of_mode_functions}} 

	\begin{figure}[t]
		\centering
		\includegraphics[width=0.48\textwidth]{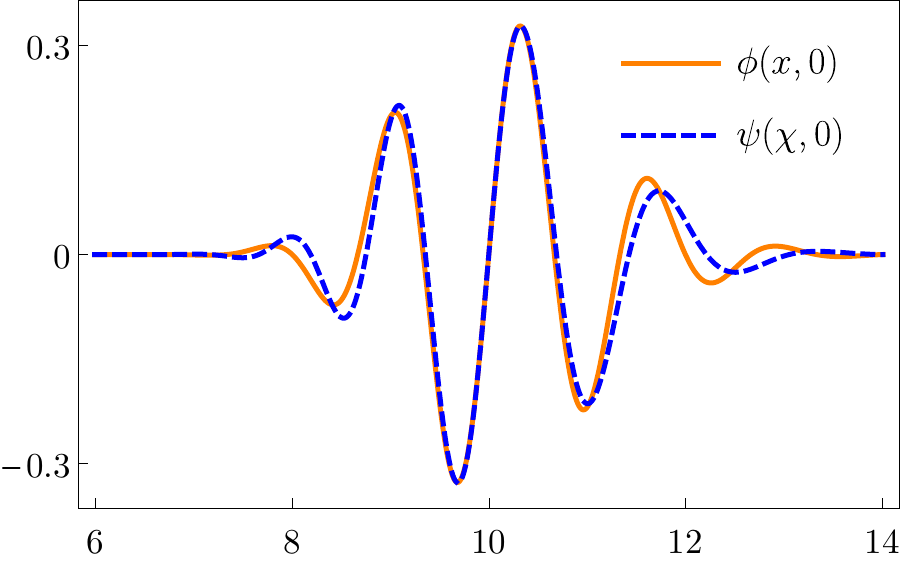}
		\caption{The initial condition for the inertial wavepacket $\phii$ and the accelerating wavepacket $\psii$.
		We obtain $\psii$ by deforming $\phii$ in the same way as the modes of a Dirichlet cavity deform under acceleration.
		The initial conditions for $\phiii$ and $\psiii$ are the same as shown here but mirrored with respect to spacetime origin.
		\label{fig:mode_functions}}
	\end{figure}

	\begin{figure}[t]
		\centering
		\includegraphics[width=0.48\textwidth]{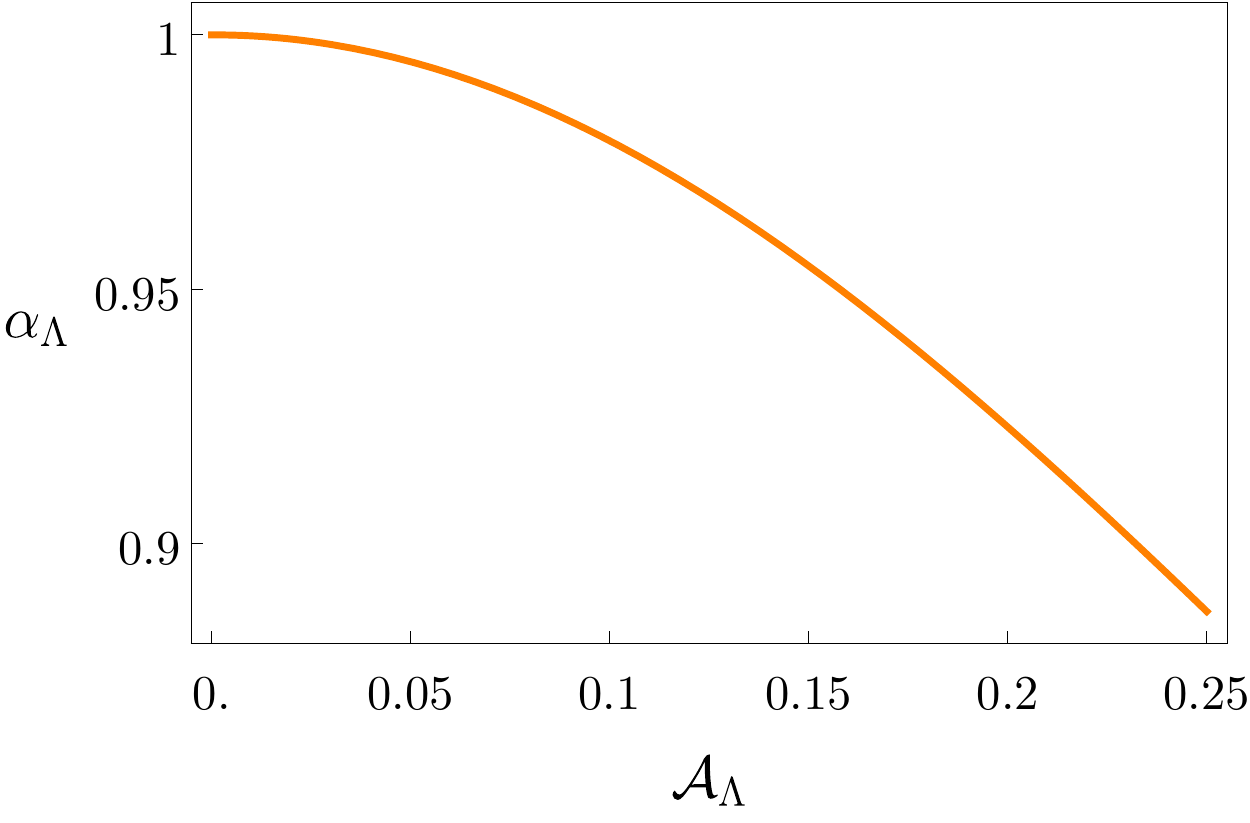}
		\caption{The dependence of the overlap $\aL = (\psiL,\phiL)$, $\Lambda \in \{\text{I},\text{II}\}$, on the proper acceleration of the wavepacket $\psiL$.
		Since we put $x_0=1/\mathcal{A}$ in Eqs.~\eqref{eq:initial_cond_phi} and~\eqref{eq:initial_cond_psi}, for $\mathcal{A} \rightarrow 0$ the initial conditions escape to infinity.
		We remedy this by interpolating between $\mathcal{A}=0$ and $\mathcal{A}=0.03$.
		\label{fig:alpha(A)}}
	\end{figure}

	We choose the mode functions $\phiL$, $\psiL$, such that their state could be prepared and measured using a finite-size apparatus.
	This means that the wave packets have to be approximately localized and have to be positive-frequency in their respective rest frames\footnote{
		Wave packets consisting of only positive-frequency plane waves cannot have compact support.
		Therefore, we allow our modes to possess infinite, but quickly vanishing tails.
	}. Furthermore, the accelerating mode functions $\psiL$ have to be far from the event horizon compared to their size $L$.
	If we denote the proper acceleration of the center of $\psiL$ by $\AL$, then this condition reads $1/\AL \gg L$.
	If this requirement is satisfied, the proper acceleration (which is a function of position in the Rindler chart) is approximately constant across $\psiL$.
	This means that we can attribute $\AL$ to $\psiL$ as their meaningful proper acceleration value.

	Similarly to Ref.~\cite{Ahmadi2016}, our choice of the mode functions satisfying the above conditions is inspired by Refs.~\cite{Doukas2013,Dragan2013a,Dragan2013}.
	The $\phiL$ are taken to satisfy the initial conditions
	\begin{subequations} \label{eq:initial_cond_phi}
	\begin{align}
		\phiL (x,0) &= \pm C e^{-2\left(\frac{x_0}{L}\log{\frac{x}{x_0}}\right)^2} \sin{\left(	\sqrt{\Omega_0^2-m^2}\left(x-x_0\right)\right)},\\
		\partial_t \phiL (x,0) &= -i \Omega_0 \phiL (x,0),
	\end{align}
	\end{subequations}
	where	$x_0$ is the position around which the function is centered, $L$ is the wave packet's width, $C$ is the normalization constant, and the upper(lower) sign corresponds to $\La = \text{I}(\text{II})$.
	$\Omega_0$, around which the spectrum of the mode function is centered, has to satisfy $\Omega_0 \gg 1/L$ so that the contribution of negative frequencies is as small as possible.
	We remove the remaining negative-frequency contribution in the numerical calculations by applying a cutoff at zero frequency.
	It can be seen that doing so leaves the spatial profile of $\phiL$ mostly intact~\cite{Ahmadi2016}.

	The choice of output mode functions $\psiL$ is independent of the choice of input mode functions $\phiL$.
	However, it is natural to obtain $\psiL$ from $\phiL$ in the same way as one obtains the modes of an accelerating cavity from the modes of a cavity at rest.
	That is, we keep the envelope but replace the trigonometric functions with modified Bessel functions and substitute the Rindler coordinates in place of inertial ones~\cite{Ahmadi2016}.
	The Rindler chart will be given by
	\begin{align}\label{rincoor}
		t&=\chi\sinh(a\eta),\nonumber \\
		x&=\chi\cosh(a\eta),
	\end{align}
	where $a$ is a positive parameter, $(x,t)$ are Minkowski coordinates, and $(\chi,\eta)$ are Rindler coordinates\footnote{
		We use the simple form of Rindler transformation with zero separation between the wedges, because our results do not depend on the minimal distance between observers (see footnote~\ref{foot:neglecting_betas}).
	}.
	The output mode functions are then given by the initial conditions
	\begin{subequations} \label{eq:initial_cond_psi}
	\begin{align}
		\psiL (\chi,0) =& C' e^{-2\left(\frac{x_0}{L}\log{\frac{\chi}{x_0}}\right)^2}  \times \nonumber\\
		& \Im\Big[I_{-i\frac{\Omega_0}{\mathcal{A}}}(m|x_0|)I_{\mathrm{i}\frac{\Omega_0}{\mathcal{A}}}(m|\chi|)\Big],\\
		\partial_{\tau} \psiL (\chi,0) =& \mp \mathrm{i} \Omega_0 \psiL (\chi,0),
	\end{align}
	\end{subequations}
	where $C'$ is a normalization constant and $I_{\mathrm{i}\nu}(x)$ is the modified Bessel function of the first kind.

	We will assume that $|x_0|=1/\mathcal{A}$, $m=0.1$, $L=2$, and $\Omega_0 \approx 5$.
	In Fig.~\ref{fig:mode_functions} we illustrate the shape of the wave packets for the above parameters and $\mathcal{A}=0.1$.

	Now that we have specified all the mode functions we can calculate the overlaps given in Eq.~\eqref{eq:overlaps}.
	We do this numerically and plot the result as a function of the proper acceleration $\mathcal{A}$ in Fig.~\ref{fig:alpha(A)}.
	We finally note that $\alii(\mathcal{A})=\ali(\mathcal{A})$, since the mode functions $\phiL$ are the same as $\phiL$ up to the reflection with respect to $x=0$.

	\subsection{Entanglement of the resource state under acceleration\label{sub:entanglement_of_the_resource_state_under_acceleration}} 

	\begin{figure}[t]
		\centering
		\includegraphics[width=0.48\textwidth]{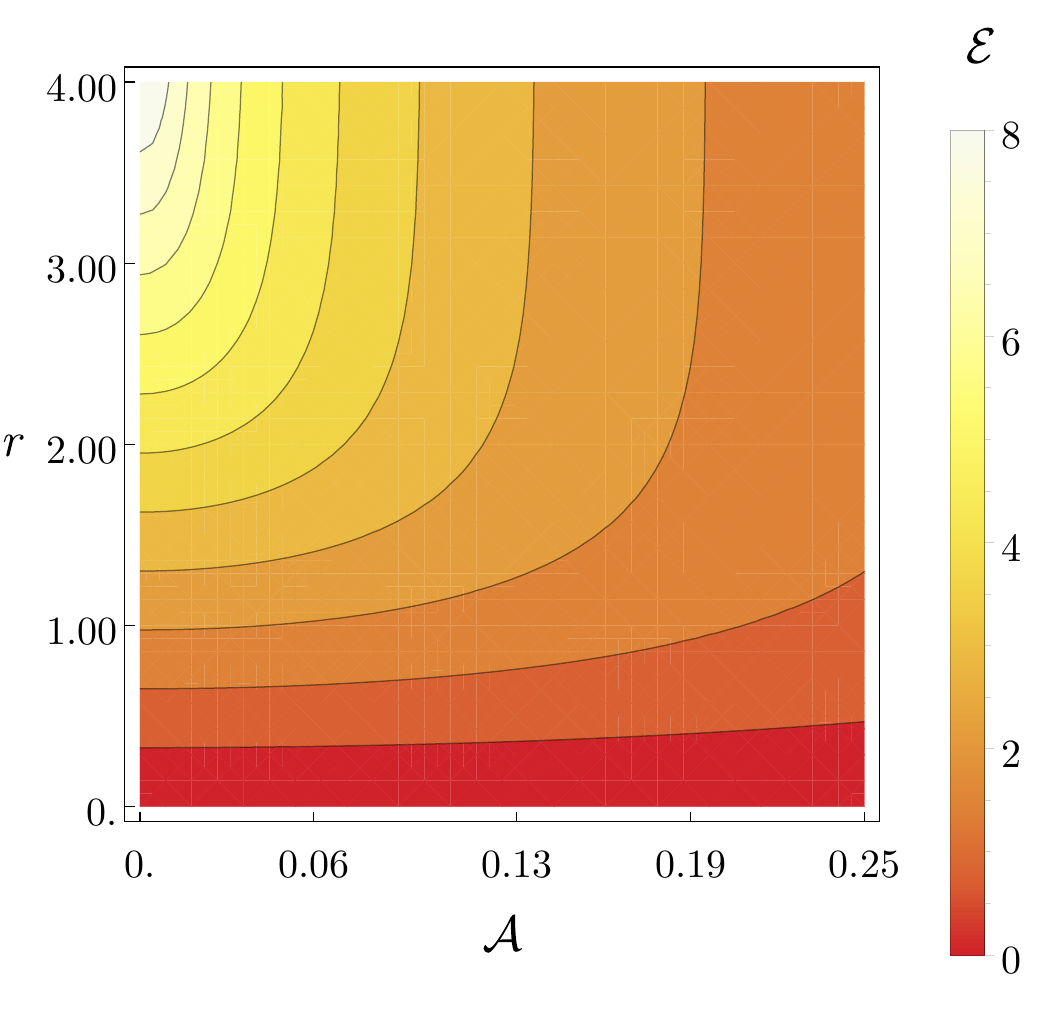}
		\caption{The logarithmic negativity $\mathcal{E}(\mathcal{A})$ of a resource state as detected by observers Alice and Bob (corresponsing to wavepackets $\psii$ and $\psiii$) moving with equal-magnitude accelerations $\Ai = \Aii = \mathcal{A}$.
		The original state is a two-mode squeezed vacuum state characterized by a squeezing coefficient $r$ of the inertial wavepackets $\phii$ and $\phiii$.
		\label{LN3}}
	\end{figure}

	\begin{figure}[t]
		\centering
		\includegraphics[width=0.48\textwidth]{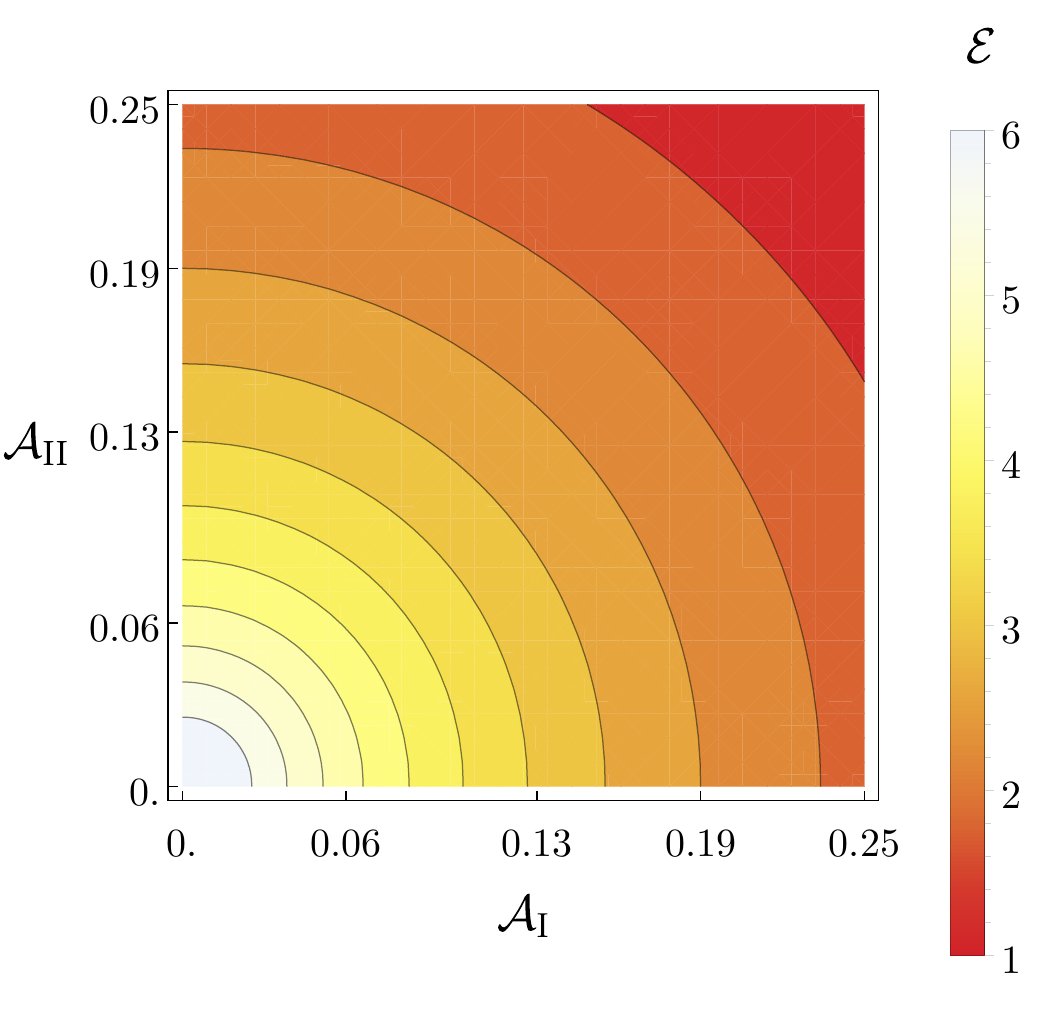}
		\caption{The logarithmic negativity $\mathcal{E}(\mathcal{A})$ of a two-mode squeezed vacuum resource state as detected by asymmetrically accelerating Alice and Bob.
		The original state is characterized by a sqeeezing coefficient $r=3.0$.
		\label{LN1}}
	\end{figure}

	Knowing $\alpha(\mathcal{A})$ we can calculate the effect of acceleration on a state prepared in an inertial frame.
	The state we consider is a two-mode squeezed vacuum state, as this is the state we will later use as a resource for the quantum teleportation and dense coding protocols.
	Without the loss of generality, we assume that the two-mode sqeezed vacuum is characterized only by the squeezing parameter~$r$.
	The covariance matrix of the state is then
	\begin{equation}
		\sigma^{(f)}=\begin{pmatrix}
		\cosh{2r} & 0 & -\sinh{2r} & 0 \\
		0  &  \cosh{2r} & 0&\sinh{2r} \\
		-\sinh{2r}  & 0 & \cosh{2r} & 0 \\
		0	 & \sinh{2r}  &0 & \cosh{2r}
		\end{pmatrix}.
	\end{equation}
	To obtain the covariance matrix $\sigma^{(d)}$ of the above state as seen by two accelerating observers, we use Eqs.~\eqref{eq:channel} and~\eqref{eq:MN_matrices}.
	The result is
	\begin{subequations} \label{eq:sigma_d}
	\begin{equation} \label{eq:sigma_d_form}
		\sigma^{(d)} = \begin{pmatrix}
		 a & 0 & -c & 0 \\
		 0  & a & 0&c \\
			-c  & 0 & b & 0 \\
		0	 & c  &0 & b
		 \end{pmatrix},
	\end{equation}
	where
	\begin{align}
		a &=  1-\ali^2 + \ali^2\cosh{2r},\\
		b &=  1-\alii^2 + \alii^2\cosh{2r},\\
		c &= \ali\alii\sinh{2r}.
	\end{align}
	\end{subequations}
	The $\aL$ are functions of the proper accelerations $\AL$ of the wavepackets $\psiL$, as illustrated in Fig.~\ref{fig:alpha(A)}.

	To understand the performance of quantum-information protocols under acceleration, it is helpful to see how the acceleration affects the entanglement of the resource state.
	We will illustrate this by comparing the logarithmic negativities of the $\sigma^{(f)}$ and $\sigma^{(d)}$ states.
	Logarithmic negativity is an entanglement monotone~\cite{Plenio2005} which is especially easy to compute for Gaussian states.
	For a state characterized by the covariance matrix of the form~\eqref{eq:sigma_d_form}, it is given by
	\begin{equation}
		{\cal E} = \max\left\{0,-\log\sqrt{\frac{\Delta-\sqrt{\Delta^2-\det \sigma^{(d)}}}{2}}\right\},
	\end{equation}
	where $\Delta = a^2 + b^2 + 2 c^2$.

	We will calculate the negativity in two different scenarios.
	The first one is a squeezed vacuum state as seen by observers moving with equal accelerations, $\Ai=\Aii=\A$.
	The logarithmic negativity of $\sigma^{(d)}$ in this case is plotted in Fig.~\ref{LN3}.
	We can see that $\mathcal{E}$ monotonically decreases with increasing acceleration and increases for larger squeezing coefficients.

	In the second scenario we allow for asymmetric accelerations, i.e. $\Ai\ne\Aii$.
	In Fig.~\ref{LN1} we now plot $\mathcal{E}$ as a function of $\Ai$, $\Aii$.
	We can observe monotonic decrease of $\mathcal{E}$ as a function of observers' accelerations.
	Moreover, the amount of entanglement increases with the increasing value of the squeezing parameter.

\section{Continuous variable quantum teleportation} 
\label{sec:continuous_variable_quantum_teleportation}

	\begin{figure}[t]
		\centering
		\includegraphics[width=0.48\textwidth]{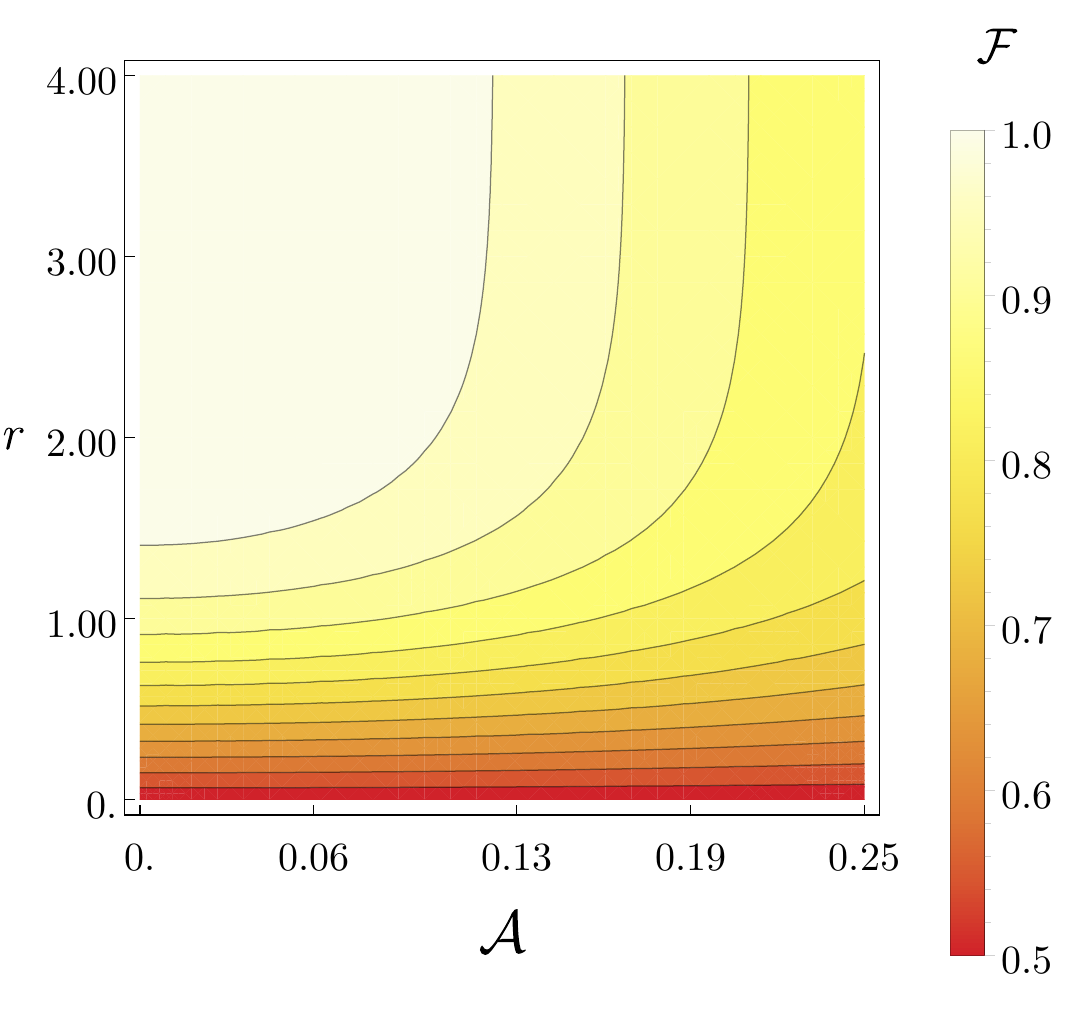}
		\caption{The fidelity $\mathcal{F}$ of quantum teleportation performed by Alice and Bob moving with equal-magnitude accelerations $\Ai = \Aii = \mathcal{A}$.
		The entangled state used as a resource is a two-mode squeezed vacuum state of $\phii$ and $\phiii$.
		$r$ is the squeezing coefficient characterizing the resource.
		\label{r12}}
	\end{figure}

	\begin{figure*}[t]
		\centering
		\includegraphics[width=1\textwidth]{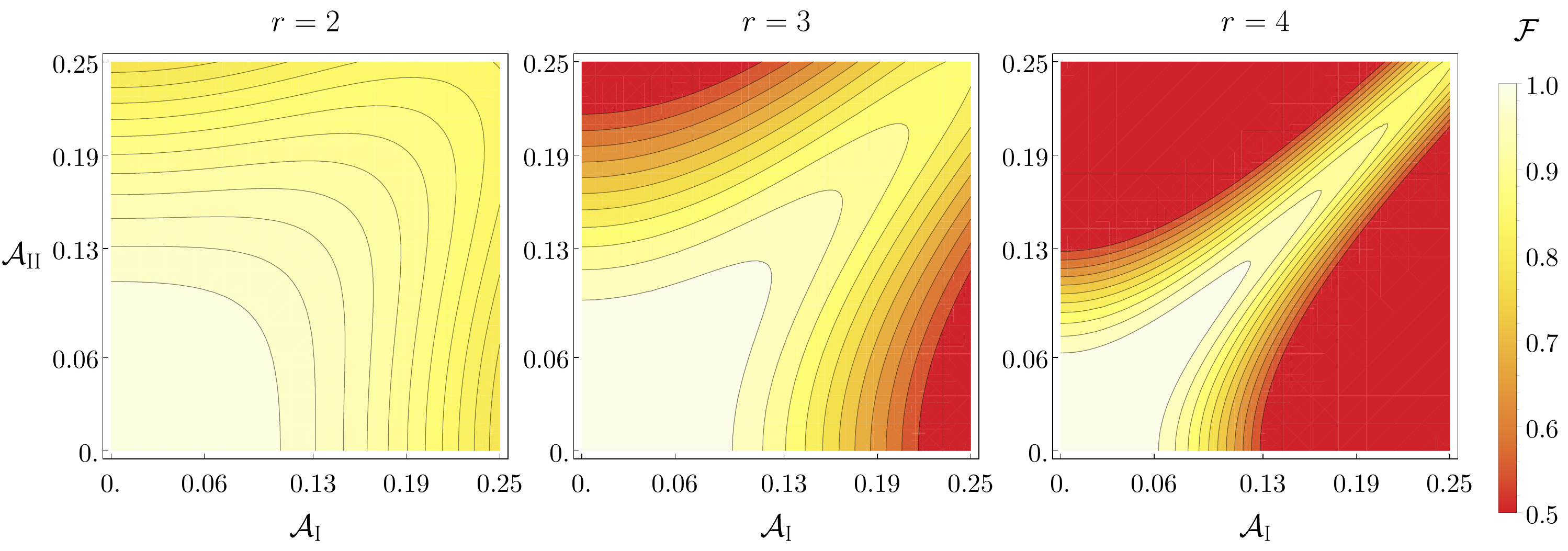}
		\caption{The teleportation fidelity $\mathcal{F}$ in the case when the magnitudes of Alice's and Bob's accelerations are independent.
		Fidelity values below $0.5$ are clipped as $\mathcal{F}=0.5$ is achievable with a classical strategy.
		$r$ is the squeezing coefficient of the two-mode squeezed vacuum resource.
		\label{r13}}
	\end{figure*}

	Continuous variable quantum teleportation (CVQT) is a major example of a quantum protocol that can utilize two-mode Gaussian states.
	Its purpose is to transfer quantum information between two spatially separated observers, Alice and Bob.
	The two parties use classical communication and the entanglement of a shared two-mode state to destroy the input state at Alice's location and reproduce it at Bob's.

	We consider the quantum teleportation protocol introduced in Ref.~\cite{Vaidman1994} and generalized in Ref.~\cite{Braunstein1998} to use resource states with non-perfect correlations.
	The protocol assumes Alice and Bob each have access to one bosonic field mode, $\psii$ and $\psiii$ respectively.
	These two modes are prepared in a Gaussian state with vanishing first moments.
	Its covariance matrix $\sigma$ can be represented in a block form
	\begin{equation}\label{eq:abc}
		\sigma=\begin{pmatrix}
			\sigma_{\text{I}} & \gamma_{\text{I,II}}  \\
			\gamma_{\text{I,II}}^T & \sigma_{\text{II}} 
		\end{pmatrix},
	\end{equation}
	where $\sigma_{\text{I}}$, $\sigma_{\text{II}}$, and $\gamma_{\text{I,II}}$ are $2\times 2$ real matrices.
	Moreover, Alice has access to one additional mode, which we call the \emph{input} mode.
	This mode is prepared in an arbitrary Gaussian state characterized by a covariance matrix $\sigma_{\text{in}}$ and (possibly nonzero) mean which Alice does not know.
	Then she performs double homodyne detection, effectively measuring two quadratures $\hat{q}_+ = \hat{q}_\text{I} + \hat{q}_{\text{in}}$ and $\hat{p}_- = -\hat{p}_\text{I} + \hat{p}_{\text{in}}$.
	As a result, she obtains a complex number, $x=\bar{q}_++i\bar{p}_-$, which is sent to Bob with the help of the classical channel.
	Bob applies a displacement $D(x)$ on mode $\psiii$ that yields a Gaussian state similar to the one originally in the input mode.

	The performance of the protocol is described by the teleportation fidelity $\mathcal{F}$, which is the overlap of the input and the output states averaged over all the possible outcomes of Alice's measurement.
	If the input state is Gaussian and pure, we have~\cite{Fiurasek2002}:
	\begin{equation}\label{eq:fid}
		\mathcal{F} = \frac{1}{\sqrt{\operatorname{det}  \Gamma}},
	\end{equation}
	where
	\begin{equation}
		\Gamma = 2 \sigma_\text{in} + \zeta\sigma_{\text{I}}\zeta + \sigma_{\text{II}} +\zeta\gamma_{\text{I,II}} + \gamma_{\text{I,II}}^T\zeta^T, \ \ \ \zeta=\operatorname{diag} (1,-1).
	\end{equation}
	This formula still holds if Alice and Bob share a state characterized by arbitrary first moments.
	However, in this case the protocol slightly changes as Bob has to perform an additional displacement~\cite{Pirandola2006}.

		Let us consider a scenario in which Alice and Bob, initially at rest, start accelerating uniformly and try to perform the CVQT protocol.
		They share a two-mode squeezed vacuum state and want to teleport an unknown coherent state that Alice has access to.
		Within the framework discussed in Sec.~\ref{sec:effect_of_acceleration_on_states}, we can study geometries in which Alice and Bob move with arbitrary accelerations, independent from each other.
		The only requirement is for Bob to be in Alice's future light-cone as it is necessary for the classical information to be sent.

		Alice and Bob's motion is accounted for by applying the two-mode channel~\eqref{eq:channel} to the resource state.
		The result, given by Eq.~\eqref{eq:sigma_d}, is characterized by two independent parameters, $\ali$ and $\alii$.
		Inserting it into Eq.~\eqref{eq:fid} gives the teleportation fidelity:
		\begin{widetext}
		\begin{equation}\label{eq:f1}
			\mathcal{F}\pqty{\ali,\alii,r}=\frac{1}{2+\frac{1}{2}\pqty{\ali^2+\alii^2}\pqty{\cosh{2r}-1} - \ali\alii\sinh{2r}}.
		\end{equation}
		\end{widetext}

		\begin{figure}[t]
			\centering
			\includegraphics[width=0.48\textwidth]{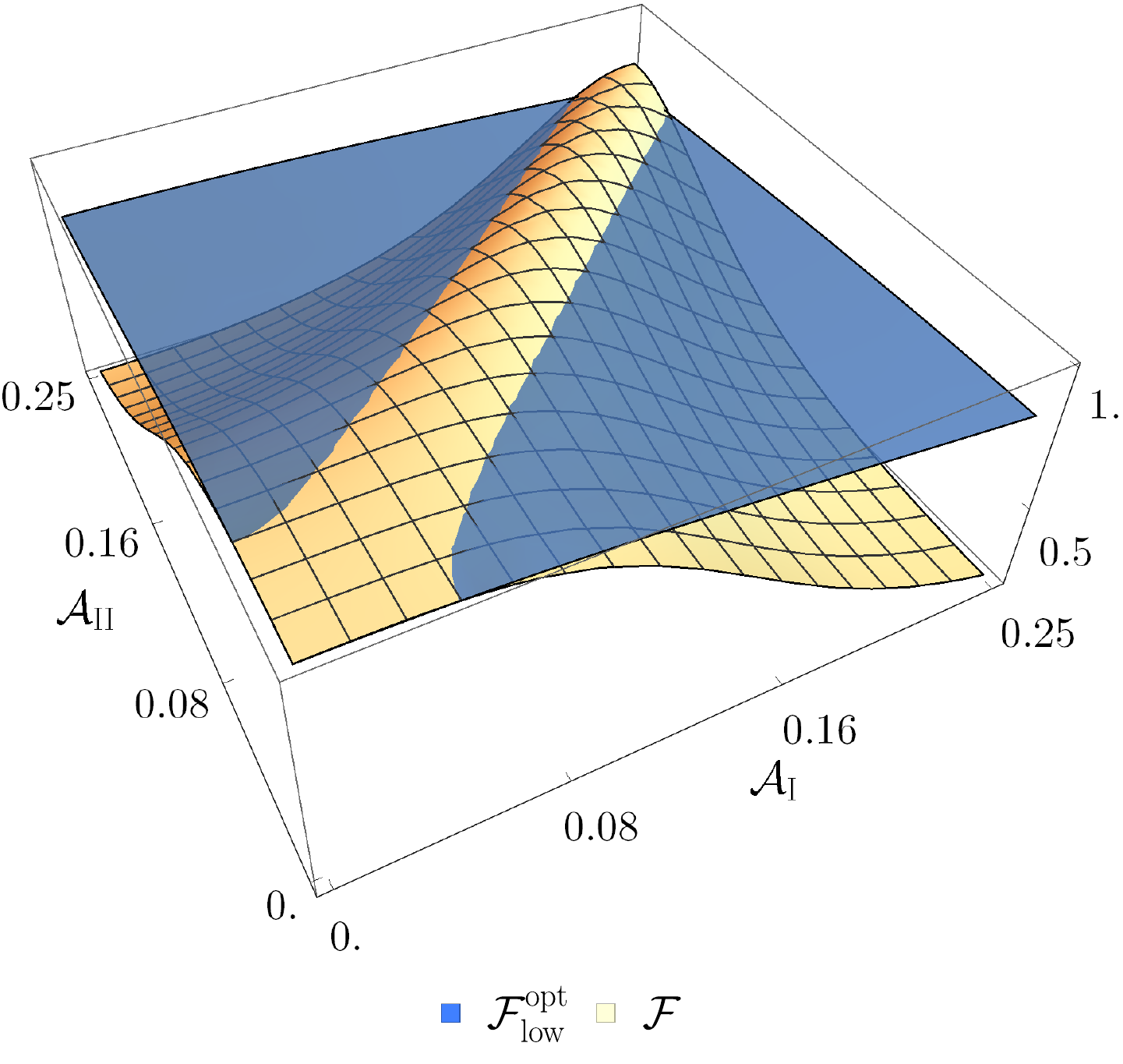}
			\caption{The teleportation fidelity of the protocol we consider (yellow, checked) and the lower bound on optimal teleportation fidelity (blue) as a function of Alice and Bob's accelerations, $\Ai$ and $\Aii$. The resource state is characterized by the squeezing parameter $r = 3.5$.
			\label{3d}}
		\end{figure}

		Firstly, we analyse the scenario in which Alice and Bob move with equal accelerations, $\Ai=\Aii=\mathcal{A}$.
		The teleportation fidelity~$\mathcal{F}(\mathcal{A})$ of the protocol they perform is plotted in Fig.~\ref{r12}.
		It decreases monotonically with the acceleration~$\mathcal{A}$, and increases for larger value of the squeezing parameter~$r$.

		In the next scenario we consider Alice and Bob moving with different accelerations, $\Ai \neq \Aii$.
		In Fig.~\ref{r13}, we plot the teleportation fidelity as a function of $\Ai$ and $\Aii$ for different values of the squeezing parameter of the resource.
		Fixing one of the accelerations, e.g. $\Ai$, we see that the teleportation fidelity as a function of the other acceleration peaks at $\Ai = \Aii$.
		This peak gets more pronounced for larger values of the squeezing parameter.

		We note that the drop of teleportation fidelity for asymmetric accelerations cannot be explained by degradation of entanglement of the resource state.
		This can be seen by looking at Fig.~\ref{LN1}, which shows that the logarithmic negativity of the resource state is essentially insensitive to asymmetry of accelerations.
		We conclude therefore that the poor performance of quantum teleportation for $\Ai \neq \Aii$ is a sign that the protocol we use is not optimal in this regime.
		We confirm this by calculating the lower bound for optimal teleportation fidelity~\cite{Mari2008}:
		\begin{equation}\label{eq:f_low_bound}
			\mathcal{F}^{opt} \ge \frac{1+\nu}{1+3\nu},
		\end{equation}
		where $\nu$ is the smallest symplectic eigenvalue of the resource state.
		Comparing this bound to the fidelity of our protocol (see Fig.~\ref{3d}) we see that the latter is strictly suboptimal for highly asymmetric accelerations.
		The asymmetry-related fidelity loss is therefore not fundamental and can be remedied, as shown in Sec.~\ref{sec:locc_optimization}, with an additional LOCC performed by Alice and Bob before the measurements.

		However, degradation of fidelity also occurs in the symmetric case (see Fiq.~\ref{r12}), for which our protocol can be shown to be optimal~\cite{Adesso2005}.
		This happens because in the accelerating frame the off-diagonal terms of the resource state [see Eq.\eqref{eq:sigma_d}] are proportional to the overlaps of inertial and accelerating wavepackets.
		These overlaps, on the other hand, are strictly less than one because each wavepacket consists of only those solutions which are positive-frequency in its respective rest frame.
		Losses stemming from this mode-mismatch are fundamental and cannot be removed by any amendments to the protocol.

\section{Continuous variable dense coding\label{sec:continuous_variable_dense_coding}} 

	\begin{figure}[t]
		\centering
		\includegraphics[width=0.48\textwidth]{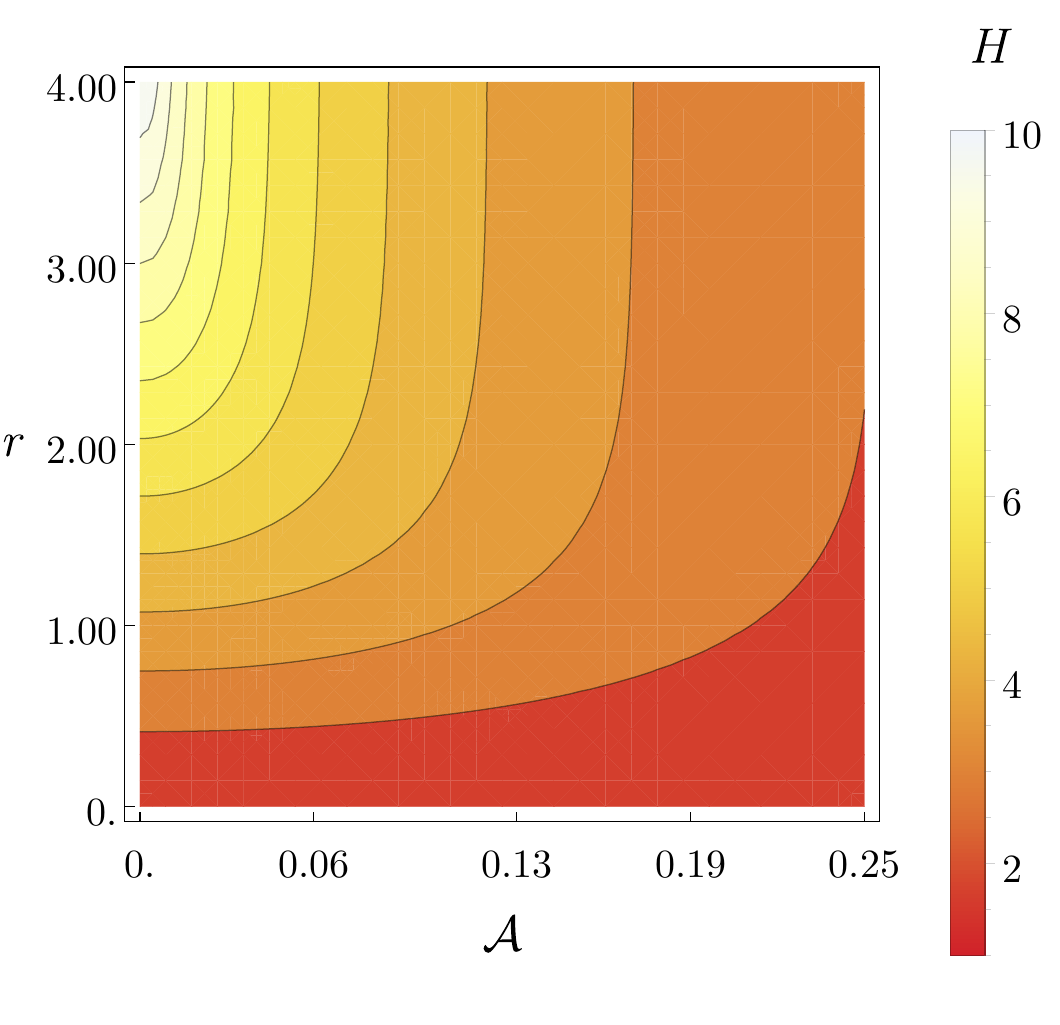}
		\caption{The mutual information $H$ of a dense coding protocol performed by Alice and Bob moving with equal-magnitude accelerations $\Ai = \Aii = \mathcal{A}$.
		$r$ is the squeezing coefficient of the resource state.
		\label{mut_inf}}
	\end{figure}

	Another quantum information protocol which can utilize two-mode Gaussian states is the continuous variable dense coding.
	Its purpose is to efficiently communicate classical information over a quantum channel between two observers, Alice and Bob.
	The two parties again share an entangled two-mode state, which allows them to communicate two real numbers while sending only one field mode.
	\begin{figure*}[hbtp!]
	\centering
	\includegraphics[width=1\textwidth]{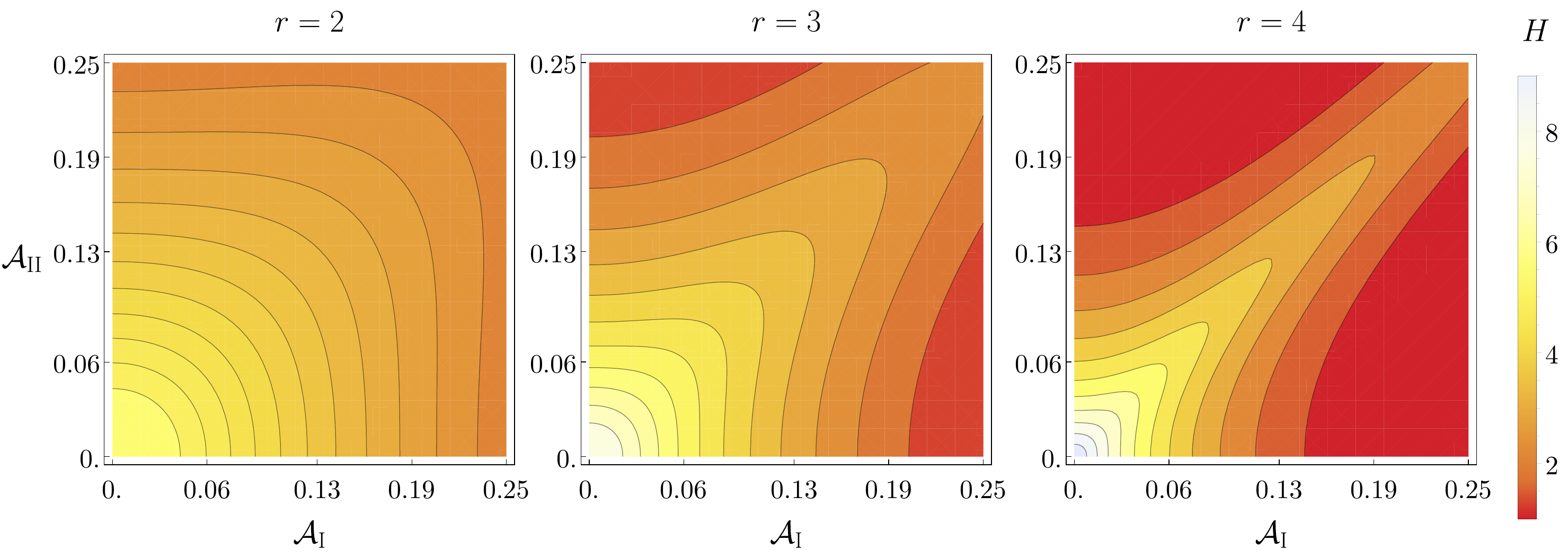}
	\caption{The mutual information $H$ for independent magnitudes of Alice's and Bob's accelerations.
		$r$ is the squeezing coefficient of the resource state.
		\label{mutual_r}}
\end{figure*}

	We consider the protocol introduced in Ref.~\cite{Braunstein2000}.
	The message that Alice attempts to communicate to Bob is a complex number $x_\text{in}$.
	The protocol assumes the message is drawn from a distribution
	\begin{equation}
		P\qty(x_\text{in}) = \frac{1}{\pi n}\exp\qty(-\frac{\abs{x_\text{in}}^2}{n}),
	\end{equation}
	where $n$ is a normalization constant.
	The protocol starts with Alice encoding the information she wants to send by performing a displacement $D\qty(x_\text{in})$ on the state of mode $\psii$, which she has access to.
	Alice then sends mode $\psii$ to Bob, who combines it with his mode $\psiii$ on a 50/50 beam splitter.
	Finally, Bob performs a homodyne detection of the resulting modes to obtain $x_\text{out}$ which is his estimate of the message that Alice has sent.

	We will use mutual information to quantify how well Bob's estimate approximates the original message.
	Mutual information is a measure of the statistical dependence between two random variables.
	It ranges from 0, when the variables are independent, to the entropy of one of them, when they are well-defined functions of each other.
	The mutual information $H$ between random variables $x_\text{in}$ and $x_\text{out}$ is defined as
	\begin{equation}
		H\qty(x_\text{in},x_\text{out}) = \int \dd{x_\text{in}}\dd{x_\text{out}}p\qty(x_\text{in},x_\text{out}) \log{\frac{p\qty(x_\text{in},x_\text{out})}{p\qty(x_\text{in})p\qty(x_\text{out})}},
	\end{equation}
	where $p\qty(x_\text{in})$, $p\qty(x_\text{out})$ are probability densities of $x_\text{in}$ and $x_\text{out}$, and $p(x_\text{in},x_\text{out})$ is a joint probability distribution.

	In Ref.~\cite{Lee2014} the mutual information for the considered protocol was calculated given an arbitrary two-mode resource:
	\begin{equation}\label{eq:res_H}
		H =\frac{1}{2} \log{ \qty[ \qty(1+\frac{n}{2V_{q_{+}}}) \qty(1+\frac{n}{2V_{p_{-}}}) ] },
	\end{equation}
	where $V_{q_{-}}$, $V_{p_{+}}$ are the variances of the quadratures $\hat{q}_+ = \hat{q}_\text{I} + \hat{q}_{\text{II}}$ and $\hat{p}_- = -\hat{p}_\text{I} + \hat{p}_{\text{II}}$, respectively.
	
	The scenario we consider now to assess the performance of dense coding is the same as for quantum teleportation.
	Alice and Bob again have access to a two-mode squeezed vacuum state as the resource, and perform the protocol under uniform acceleration.
	The influence of their motion is once again described by applying the two-mode channel~\eqref{eq:channel} to the resource state.
	By inserting~\eqref{eq:sigma_d} into~\eqref{eq:res_H}, we arrive at the expression for mutual information between Alice's message and Bob's estimate:
	\begin{widetext}
	\begin{equation}
		H =  \log{\qty(1 +\frac{n}{2 + \qty(\ali^2+\alii^2)\qty(\cosh{2r}-1) - 2 \ali \alii \sinh{2r}} ) }.
	\end{equation}
	\end{widetext}

	Firstly, we analyze the case when Alice and Bob accelerate with the equal magnitudes $\Ai=\Aii=\mathcal{A}$.
	The mutual information $H(\mathcal{A})$ of the protocol in this situation is plotted in Fig.~\ref{mut_inf}.
	It decreases for larger values of the acceleration and increases with the squeezing parameter in a similar way as the teleportation fidelity in Fig.~\ref{r12}.

	The next scenario is the asymmetric case, $\Ai \ne \Aii$.
	In Fig.~\ref{mutual_r} we plot the mutual information as a function of $\Ai$ and $\Aii$ for different values of the squeezing parameter $r$.
	Again, the mutual information behaves similarly to the teleportation fidelity (see Fig.~\ref{r13}), achieving a distinctive peak at $\Ai =\Aii$.

	The similarity of the results for dense coding and quantum teleportation suggests that the causes of the efficiency loss are the same in both cases.
	Entanglement degradation due to the acceleration again plays a fundamental role and leads to the reduction of the mutual information between the observers in every scenario.
	Moreover, as the drop of the efficiency in the asymmetric setup cannot be explained by entanglement degradation alone, we conclude that the protocol is not well suited for the resource in this case.
	However, as shown in the following section, the protocol can be improved by adding an extra, motion-dependent LOCC operation.

\section{Reducing the effect of asymmetry with LOCC\label{sec:locc_optimization}} 
	We now proceed to characterize a local Gaussian map which improves the performance of quantum teleportation and dense coding for asymmetric resource states.
	At the level of the covariance matrix, a general TGCP (trace-preserving, Gaussian, and completely positive) map acts as
	\begin{equation}
		\sigma \rightarrow \sigma' = S \sigma S^T + G,
	\end{equation}
	where $S$ corresponds to the unitary operation and $G$ to the added noise.
	Since we consider a local channel, $S=S_{\ii} \oplus S_{\iii}$ and $G=G_{\ii} \oplus G_{\iii}$.
	We will take~\cite{Mari2008}
	\begin{subequations}
	\begin{align}
		S_{\ii} &=
		\begin{cases} 
			\tan{\theta} \ \openone & 0 < \theta \leq \pi/4 \\
			\openone & \pi/4 \leq \theta < \pi/2,
		\end{cases}\\
		S_{\iii} &=
		\begin{cases}
			\openone & 0 < \theta \leq \pi/4 \\
			\cot{\theta} \ \openone & \pi/4 \leq \theta < \pi/2,
		\end{cases}\\	
		G_{\ii} &=
		\begin{cases}
			(1-\tan^2{\theta}) \openone & 0 < \theta \leq \pi/4 \\
			0 & \pi/4 \leq \theta < \pi/2 ,
		\end{cases}\\
		G_{\iii} &=
		\begin{cases}
			0 & 0 < \theta \leq \pi/4 \\
			(1-\cot^2{\theta}) \openone & \pi/4 \leq \theta < \pi/2,
		\end{cases}
	\end{align}\label{eq:opt_chan}
	\end{subequations}
	where
	\begin{subequations}
	\begin{equation}
	\theta=\arctan{\sqrt{\frac{1-\epsilon}{1+\epsilon}}},
	\end{equation}
	\begin{equation}
		\epsilon = \frac{\sqrt{2}(\ali^2-\alii^2)\sinh{r}}{\sqrt{(\ali^2+\alii^2)^2 (\cosh{2r}-1)+8\ali^2\alii^2}}.
	\end{equation}
	\end{subequations}
	Once again $r$ is the squeezing parameter of the resource state and $\ali$, $\alii$ are the overlaps introduced in Eq. \eqref{eq:overlaps}.
	
\begin{figure}[t]
		\centering
		\includegraphics[width=0.48\textwidth]{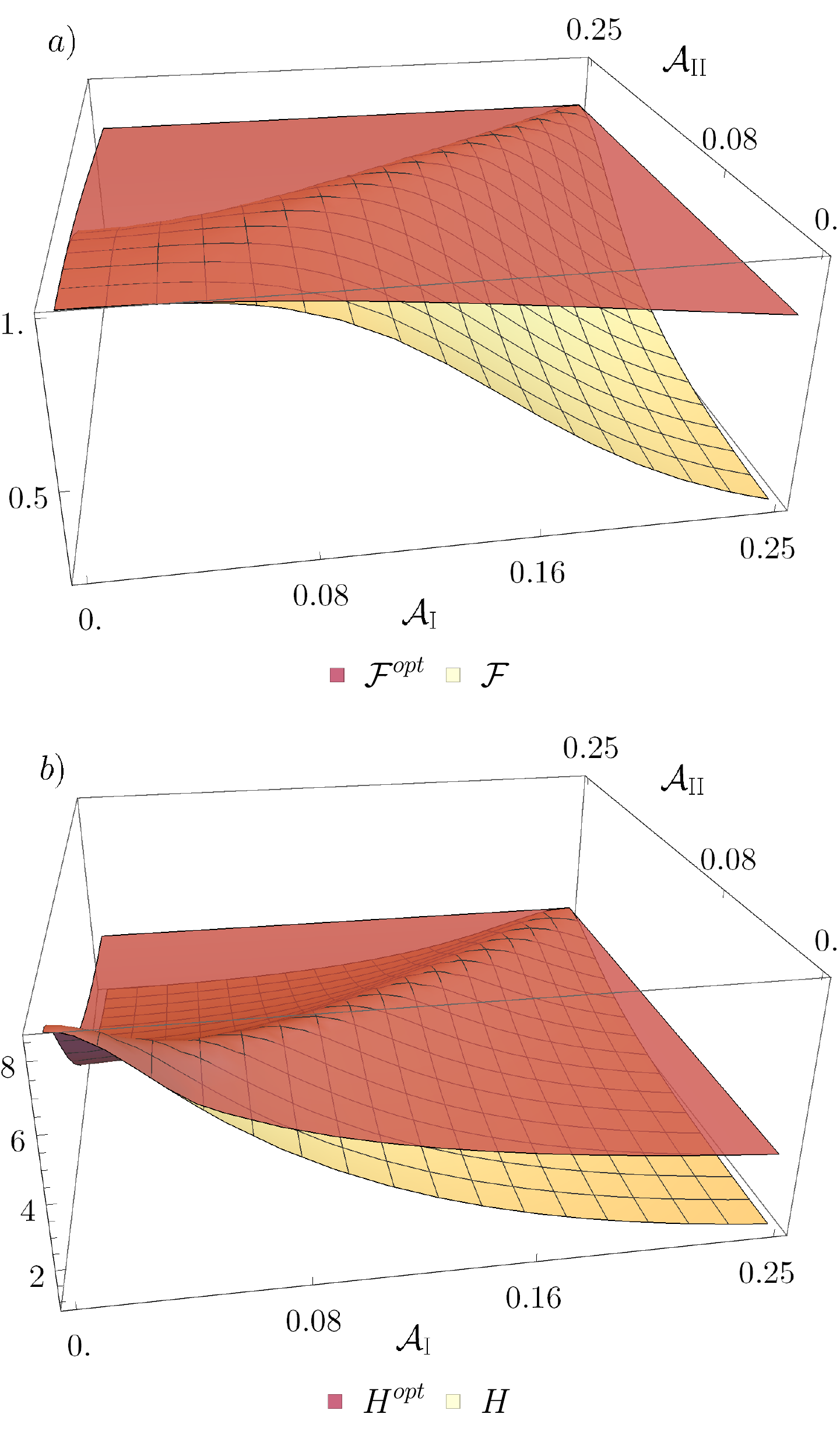}
		\caption{(a) The teleportation fidelity without (yellow, checked) and with (red) the asymmetry-compensating LOCC.
		(b) Mutual information of the dense coding protocol without (yellow, checked) and with (red) asymmetry compensation.
		In both cases the losses due to unequal accelerations of the observers are almost completely removed.
		The resource state is characterized by the squeezing parameter $r = 3.5$.\label{optim}}
	\end{figure}

	In the quantum teleportation protocol, the channel~\eqref{eq:opt_chan} should be applied to the resource state just before the measurements.
	The resulting fidelity can be calculated from Eq.~\eqref{eq:fid}, becoming
	\begin{equation} \label{eq:optimized_fidelity}
		\mathcal{F_\text{opt}}=\frac{1+|\epsilon|}{1+\nu+2 |\epsilon|},
	\end{equation}
	where $\nu$ is the smallest symplectic eigenvalue of the partially transposed resource state before the channel~\eqref{eq:opt_chan} is applied.
	It may happen that the above fidelity is lower than the initial one, especially when the resource state is symmetric enough from the beginning.
	For asymmetric states, however, the improvement is significant (see Fig.~\ref{optim}(a)).
	In fact, Ref.~\cite{Mari2008} proves that~\eqref{eq:optimized_fidelity} always exceeds the lower bound~\eqref{eq:f_low_bound}.
	This implies that the fidelity cannot be further improved by more than $0.086$, which is the maximum difference between~\eqref{eq:f_low_bound} and the upper bound for fidelity.

	We show now that the channel~\eqref{eq:opt_chan} also improves the performance of dense coding.
	This time, it is applied to the resource state before Bob sends his mode to Alice.
	The mutual information~\eqref{eq:res_H} then becomes
	\begin{equation}
		H^\text{opt} = \log\qty(1+\frac{n}{2}\frac{\mathcal{F}^\text{opt}}{1-\mathcal{F}^\text{opt}}),
	\end{equation}
	where $\mathcal{F}^\text{opt}$ is given by Eq.~\eqref{eq:optimized_fidelity}.
	To see how well the effect of asymmetry is removed, for each asymmetrically accelerated resource state we will use the asymmetric resource with equal entanglement as a reference.
	We find that the difference of mutual information values for those two cases is always smaller than $15\%$, for $n=10$, and decreases with $n$.
	The comparison is shown in Fig.~\ref{optim}(b).

	Finally, we note that the channel~\eqref{eq:opt_chan} can be decomposed into a (Gaussian) local unitary, followed by an attenuation performed locally either by Alice or Bob.
	This provides a clear operational interpretation of the asymmetry-compensation step.
	Firstly, Alice and Bob both have to perform exact combination of phase space rotations and squeezings of their respective modes and then, one of them has to use a beam splitter with one unused port and a given transmissivity.

\section{Conclusions\label{sec:conclusions}}
	In this article, we have studied the effect of relativistic acceleration on continuous variable quantum information protocols, applying the results of Ref.~\cite{Ahmadi2016}. 
	Within the framework introduced there, we have been able to compute how two uniformly accelerating observers detect the state of two inertial, bosonic, approximately localized wave-packets.
	Such a description can be represented as the action of a Gaussian quantum channel, which allowed us to efficiently calculate covariance matrix of the resource state in the Rindler frame of reference.

	We have assumed that a pair of observers, Alice and Bob, have access to the two-mode squeezed vacuum state prepared in the inertial frame.
	They accelerate and perform a quantum information protocol.
	We have considered different scenarios.
	Observers can counter- or co-accelerate with different magnitudes and have an adjustable spatial separation.
	To quantify the effectiveness of the protocol, we have used the teleportation fidelity in the case of the quantum teleportation and the mutual information for the dense coding.

	We have identified two types of efficiency losses that are present under the acceleration.
	Firstly, the decompositions of the free field in Minkowski and Rindler spacetimes into positive and negative components are different.
	It makes the construction of the positive-frequency wave-packets that take the same form in both frames impossible.
	In result, the efficiency of the protocols drops as it explicitly depends on the overlaps of inertial and non-inertial mode functions.
	Secondly, if Alice's and Bob's accelerations are different, the resource state becomes asymmetric and thus ill-suited for the standard protocols we consider.
	We have shown, however, that the asymmetry losses can be reduced if Alice and Bob perform additional local adjustments prior to measurements.
	In particular, we demonstrated that two Gaussian unitaries followed by attenuation of one of the modes can recover at least $89\%$ of maximal teleportation fidelity and $85\%$ of the maximal mutual information.
	For quantum teleportation, a small further improvement with Gaussian operations is still possible~\cite{Mari2008}.
	For dense coding, however, full optimization of the resource state remains an open problem.

	Regarding the outlook, we have managed to calculate the effect of relativistic acceleration on quantum information protocol in a realistic setup.
	It further proves that framework we have applied can be readily used to study an effect of acceleration on any type of quantum-information procedure involving Gaussian states.
	Further work might include the analysis of other relevant quantum information protocols, such as quantum key distribution or quantum bit commitment.

\section*{Acknowledgments}
This work was supported by National Science Centre, Sonata BIS Grant No. DEC-2012/07/E/ST2/01402.

\end{document}